# Movable Antenna Enabled Reconfigurable Array Topologies for Structured Beam Communications

Hongyun Jin, *Student Member, IEEE*, Wenchi Cheng, *Senior Member, IEEE*, and Jingqing Wang *Member, IEEE*

*Abstract*—Spatially structured beams have emerged as a promising technology for enhancing spectrum efficiency (SE) in sixth-generation (6G) networks. However, structured beam schemes based on fixed-position antennas (FPAs) fail to fully exploit the array aperture, thereby limiting their topological reconfigurability and adaptability to diverse communication scenarios. To overcome these limitations, this paper proposes a novel structured beam communication framework exploiting movable antennas (MAs) to achieve reconfigurable array topologies. Specifically, we develop an MA-based geometric modeling framework to construct a variety of practical array topologies, thereby enabling the realization of diverse array configurations utilizing a unified hardware platform. Furthermore, we investigate the joint design of the array topology and the structured beamforming vector to efficiently exploit the array aperture and facilitate the multiplexing of orthogonal spatial modes. Accordingly, we formulate the corresponding beam generation and demodulation schemes and derive the channel gains under varying array topologies. We also propose an alternating optimization algorithm to jointly optimize the array topology configuration, the antenna element positions, and the structured beamforming vector, with the aim of maximizing the system SE. Numerical results demonstrate that the proposed joint design significantly enhances the SE compared to conventional FPA schemes. By synergizing the spatial multiplexing degrees of freedom (DoFs) of structured beams with the mobility DoFs of MAs within 2D planar regions, this work establishes a reconfigurable and practical framework for structured beam-based wireless communications.

## I. INTRODUCTION

**W**ITH the increasing demand for higher data rates and more efficient communication systems in sixth-generation (6G) networks, the development of novel techniques for improving spectrum efficiency (SE) has become a critical research challenge [1]. Spatially structured beams have emerged as a promising technique for wireless communications due to their potential to enhance communication capacity [2], [3]. Structured electromagnetic beams encompass a class of beams characterized by spatially structured distributions of amplitude, phase, and polarization. The synthesis of such waves fundamentally relies on the manipulation of electromagnetic eigenmodes [4]. Specifically, through the coherent superposition of multiple orthogonal modes, it becomes

This work was supported in part by the National Natural Science Foundation of China (No. 62341132) and in part by QTZX24078. *(Corresponding author: Wenchi Cheng.)*

H. Jin, W. Cheng, and J. Wang are with the State Key Laboratory of Integrated Services Networks, Xidian University, Xi'an, 710071, China (e-mails: hongyunjin@stu.xidian.edu.cn, wccheng@xidian.edu.cn, and jqwangxd@xidian.edu.cn).

feasible to construct beam patterns exhibiting intricate spatial phase and amplitude profiles [5]. As a representative paradigm of structured beams, electromagnetic waves carrying orbital angular momentum (OAM) exhibit a helical phase front. This unique structural characteristic provides an additional degree of freedom, enabling the multiplexing of multiple independent communication channels within the same frequency band [6]. Consequently, structured beam transmission is capable of significantly improving SE, making it well-suited for high-capacity wireless communication systems. However, despite this promising potential, practical implementations utilizing OAM suffer from severe beam divergence in higher-order modes during long-distance propagation, which significantly degrades SE. This limitation motivates the joint design of reconfigurable array topology and beamforming vectors to enable high-capacity structured beam communication.

Recent advancements in antenna array technologies have led to the development of various array architectures, which offer performance gains in specific communication scenarios. The authors of [7] proposed a taxonomy framework for multiple-input multiple-output (MIMO) and antenna array technologies, classifying twelve representative approaches into three major categories and discussing their potential benefits, challenges, and key enabling factors. Regarding structured beam synthesis, the authors of [8] proposed an $N$-dimensional quasi-fractal uniform circular array (ND QF-UCA) antenna structure and developed corresponding $N$-dimensional structured beams generation and demodulation schemes for mode-division multiplexing. This approach enables the utilization of more orthogonal spatial modes than the number of array elements, thereby achieving higher SE. The authors of [9] proposed a pre-optimized irregular array (PIA) that optimizes the base station antenna locations to maximize the average sum rate over a given coverage area, and demonstrated significant sum-rate gains compared with conventional compact and sparse uniform arrays. The authors of [10] proposed a uniform linear array (ULA) design principle that constructs sparse arrays by concatenating multiple sub-ULAs, thereby enabling sparse-array design through a polynomial model while achieving a large uniform degree of freedom with low mutual coupling. The authors of [11] introduced an OAM-based structured beam transmission framework based on a continuous-aperture array realized by a hybrid antenna integrating fluid antennas and fixed-position antennas (FPAs), and derived an electromagnetic-channel model together with an upper bound on the SE from an electromagnetic information theory perspective.

Movable antenna (MA) systems have attracted considerable attention as an emerging reconfigurable technology, in which antenna elements dynamically adjust their positions to exploit location diversity and adapt to instantaneous propagation conditions [12]–[14]. As a result, MA systems can achieve higher array gains and improved communication performance compared with conventional static arrays, thereby opening new opportunities for future wireless systems. The authors of [15] proposed several general MA architectures for wireless communications by extending existing designs along key dimensions to support different application scenarios and cost constraints, and further surveyed candidate implementation approaches based on direct mechanical or equivalent electronic control. The authors of [16] proposed a novel MA architecture and developed a field-response multipath channel model, based on which the maximum achievable MA channel gain was characterized under both deterministic and stochastic channels, demonstrating significant performance improvements over FPAs. The authors of [17] investigated MA-enabled multigroup multicast transmission and jointly optimized the MA positions and transmit beamforming to maximize the minimum weighted SINR, demonstrating significant communication performance gains over FPAs and schemes employing only partial MA deployment. The authors of [18] presented a MA communication system prototype with ultra-precise movement control to experimentally verify the performance gains of MA in practical environments. Extensive experimental results demonstrated the significant potential of MA technology for enhancing wireless communication performance.

For high-capacity structured beam transmission, array-based communication fundamentally relies on the joint design of the antenna array and structured beams, which demands the joint optimization of the array topology and modulation vectors. However, conventional arrays with FPAs lack the flexibility to reconfigure their topologies [19], thereby limiting their adaptability to diverse communication scenarios. Despite this limitation, the potential of MAs to enhance structured beam communication remains largely untapped in existing literature. The inherent flexibility of MAs enables the realization of diverse array topologies, facilitating the generation of structured beams with distinct gain profiles. This characteristic makes MAs intrinsically well-suited for spatially structured beam communications, underscoring the fundamental importance of investigating MA-enabled reconfigurable array architectures. Nevertheless, for existing MA-based systems, the computational complexity associated with antenna position optimization and beamforming design remains a significant challenge for practical implementation. Furthermore, most existing studies on MA systems have not sufficiently investigated the design of specific array topologies for real-world wireless communication scenarios. Meanwhile, with the evolution of large-scale array communications, the increasing number of antenna elements leads to increased hardware costs and power consumption. Particularly for conventional UCA-based structured beam systems, simply increasing the element count merely yields severely divergent higher-order spatial modes, failing to generate effective structured beams and limiting the capacity potential. Consequently, achieving more efficient utilization of the array aperture, along with the exploitation of additional spatial DoFs, has become imperative [20].

These challenges strongly motivate the investigation of MA-enabled reconfigurable array topologies for structured beam communications, leveraging the unique spatial DoFs inherent in structured beams and the antenna mobility DoFs of MAs.

In order to solve the above problems, we propose a MA-enabled reconfigurable array architecture for structured beam communications. The main contributions of this paper are summarized as follows:

- First, we develop a geometric modeling framework based on MAs, from which a variety of practical array topologies are derived. Generalizing from the conventional uniform circular array (UCA), we construct a diverse set of array configurations, including concentric uniform circular arrays (CUCAs), fractal uniform circular arrays (FUCAs), uniform rectangular arrays (URAs), radial linear arrays (RLAs), and spiral arrays [21], [22]. These practical and feasible topologies can be dynamically reconfigured via antenna mobility to adapt to diverse wireless communication scenarios. Topology reconfiguration provides an efficient method to fully leverage the mobility of MAs, facilitating the flexible and rapid deployment of multiple predefined array structures. Consequently, the proposed framework serves as a promising solution for the practical implementation of MA-enabled wireless communication systems.

- Second, we investigate the joint design of the array topology and the structured beamforming vector to efficiently exploit the array aperture and facilitate the multiplexing of orthogonal spatial modes. Specifically, we develop beam generation and demodulation schemes tailored to various array topologies and analytically derive the corresponding channel gains for distinct spatial modes. Through this analysis, we demonstrate that MA-enabled arrays are intrinsically well-suited for structured beam communications, given that structured beam transmission fundamentally requires the joint design of array geometry and modulation vectors. Building on this insight, we establish a novel framework for structured beam transmission utilizing reconfigurable array topologies. Consequently, by synergizing the spatial multiplexing degrees of freedom (DoFs) of structured beams with the mobility DoFs of MAs within two-dimensional (2D) planar regions, we unlock the full potential of structured beam-based wireless communications for high-capacity transmission.

- Third, we propose an efficient alternating optimization algorithm to jointly optimize the array topology and structured beamforming vectors. We conduct extensive performance evaluations and simulation comparisons across diverse setups, encompassing various array topologies, numbers of MAs, element placements, array radii, and spatial mode multiplexing schemes. Our results systematically demonstrate the superiority of the proposed joint framework over conventional fixed-topology schemes. Furthermore, we provide theoretical insights to guide MA positioning strategies and communication scheme design in practical systems. Overall, this work establishes a feasible and scalable pathway for MA-enabled structured beam systems, facilitating large-scale, reconfigurable deployments, thereby realizing efficient, high-capacity wireless transmission.

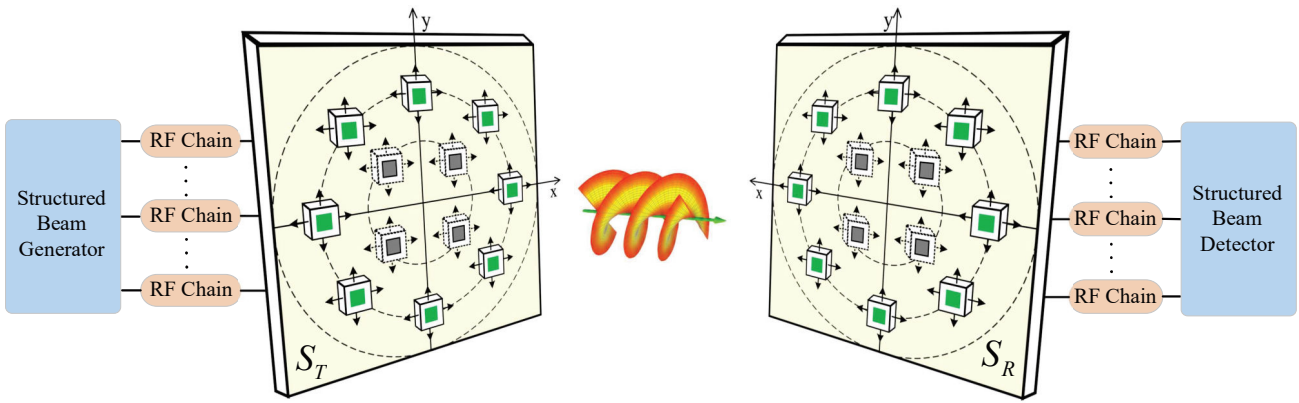

Fig. 1. The system model of movable antenna enabled reconfigurable array topologies for structured beam communications.

The rest of this paper is organized as follows. Section II introduces the system model for structured beam communication utilizing MA-enabled flexible array topologies. Section III details the geometric models based on CUCAs and FUCAs, along with their corresponding beam transmission schemes, and subsequently proposes an alternating optimization algorithm to investigate the optimal topology and achievable SE. Section V presents the numerical results and performance evaluations. Finally, Section VI concludes the paper.

*Notation:* Matrices and vectors are denoted by bold uppercase and lowercase letters, respectively. The notations $(\cdot)^T$, $(\cdot)^H$, and $(\cdot)^{-1}$ denote the transpose, conjugate transpose, and inverse, respectively. We denoted by $\mathbb{C}^{(M) \times (N)}$ the dimension of the matrix with $M$ rows and $N$ columns.

## II. SYSTEM MODEL OF MOVABLE ANTENNA ENABLED RECONFIGURABLE ARRAY TOPOLOGIES

Figure 1 illustrates the system model of the MA-enabled reconfigurable array topologies for structured beam communications. The transmitter comprises a structured beam generator, radio frequency (RF) chains, and an array composed of movable and switchable antennas. Specifically, each antenna element is connected to the RF chain via flexible cables and can be flexibly positioned within the continuous 2D region, thereby enabling the formation of arrays with arbitrary topologies. Furthermore, the inclusion of switchable antennas facilitates the dynamic control of RF activation states, significantly enhancing the reconfigurability of the array. Specifically, to synthesize specific array topologies, these antennas can be selectively activated or deactivated to form the desired array configuration, thereby preserving the integrity of the structured beam phase wavefronts. At the receiver, the system features a structured beam detector, RF chains, and a counterpart array employing movable and switchable antennas. The receiving array adopts an identical topology to that of the transmit array and utilizes the corresponding decomposition scheme to recover the orthogonal data streams. Let $\mathbf{t}_n = [x_{t,n}, y_{t,n}, 0]^T \in S_T$ and $\mathbf{r}_n = [x_{r,n}, y_{r,n}, z_{r,n}]^T \in S_R$ denote the coordinate vectors of the $n$-th MA at the transmitter and receiver, respectively, where $S_T$ and $S_R$ represent the feasible moving regions. The transmit and receive arrays comprise $N_t$ and $N_r$ antenna elements, respectively. The system is equipped with $N_f$ RF chains, satisfying the constraint $N_f \leq N_t = N_r$. In this paper, we primarily investigate the line-of-sight (LoS) scenario, assuming that the transmit and receive arrays are coaxially aligned.

In practical implementation scenarios, constrained by hardware complexity and cost, the movement of each MA is typically confined to predefined trajectories, such as rectilinear paths in Cartesian coordinates or radial and circumferential paths in polar coordinates. In this work, we focus on the synthesis of orthogonal structured beams, fundamentally utilizing the UCA structure. Given that diverse array configurations can be reconfigured from the UCA via MAs, the constructed topologies must strictly adhere to specific geometric relationships to support beam generation. Specifically, a unified hardware platform is utilized to adapt to diverse communication scenarios, obviating the need for physical antenna replacement. By leveraging software-defined reconfiguration, versatile array topologies and beam structures can be synthesized to align with specific communication scenarios. In distinct contrast to conventional massive MIMO systems that rely on scaling up the number of antenna elements, the proposed topology reconfiguration paradigm represents a strategic exploitation of the spatial DoFs inherent in MAs [23].

By integrating reconfigurable array topologies with digital beamforming schemes, we investigate the distinct advantages offered by various geometric configurations and quantify the resulting performance enhancements in structured beam transmission. To this end, we propose a holistic joint design framework that bridges the perspectives of digital signal processing and physical array geometry, aiming to realize a highly flexible and efficient transmission. Specifically, the deployment of movable and switchable antennas facilitates dynamic geometry reconfiguration. By jointly adjusting antenna positions and switching states, unprecedented flexibility in shaping the electromagnetic response of the array can be achieved, thereby supporting diverse structured beam synthesis.

## III. Geometric Modeling and Structured Beam Synthesis Framework

### A. Structured Beam Generation via Concentric UCAs

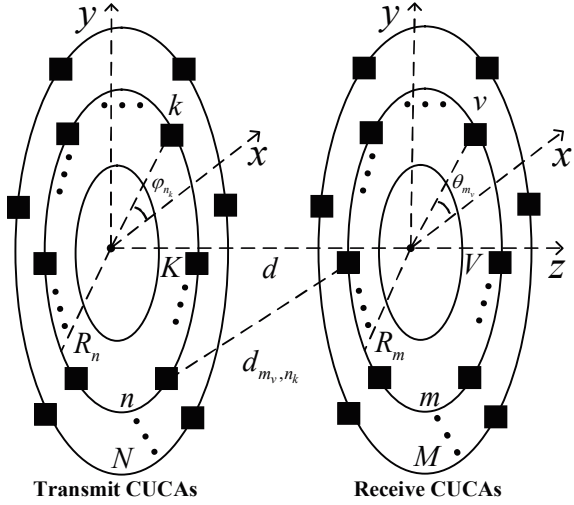

Fig. 2. Geometric configuration of the CUCAs topology.

Figure 2 depicts the geometric configuration of the CUCAs topology. The system employs a CUCAs at both the transmitter and receiver. The transmitter consists of $N$ concentric rings with radii $R_n$ ($n = 1, \ldots, N$), while the receiver comprises $M$ rings with radii $R_m$ ($m = 1, \ldots, M$). We denote by $R_{tE}$ and $R_{rE}$ the radii of the entire transmit and receive CUCAs, respectively, where $R_{tE} = R_N$ and $R_{rE} = R_M$. Generally, we assume that $R_{tE} = R_{rE}$. The transmit and receive CUCAs are assumed to be coaxially aligned and parallel, separated by a transmission distance $d$. Each ring at the transmitter is equipped with $K$ antenna elements uniformly distributed along the circumference, while each ring at the receiver comprises $V$ elements. The angular position of an element is defined as its azimuth angle with respect to the $x$-axis. Let $\varphi_{n,k}$ denote the azimuth of the $k$-th element ($k = 1, \ldots, K$) on the $n$-th transmit ring. Similarly, the azimuth of the $v$-th element ($v = 1, \ldots, V$) on the $m$-th receive ring is denoted by $\theta_{m,v}$.

Considering the uniform distribution of array-elements in CUCAs, specific criteria outlined in Eq. (1) must be adhered to concerning the array geometry of CUCA antennas. Designing the CUCA, the radii $R_n$, azimuth $\varphi_{n,k}$, and the number of array-elements in each UCA $K_n$ are required to satisfy certain predefined conditions. The conditions corresponding to CUCA can be given as follows:

$$
\text{CUCA:} \begin{cases} R_n < R_N, R_n \neq R_{n+1}, n \in [1, N); & \text{(1a)} \\ \varphi_{n,k} = \dfrac{2\pi k}{K_n}; & \text{(1b)} \\ x_{n,k} = R_n \cos \varphi_{n,k}; & \text{(1c)} \\ y_{n,k} = R_n \sin \varphi_{n,k}; & \text{(1d)} \\ \|\boldsymbol{t}_{n,k} - \boldsymbol{t}_{n+1,k+1}\| \gg D. & \text{(1e)} \end{cases}
$$

Constraint (1a) stipulates that each sub-array possesses a distinct radius $R_n$, bounded by the maximum CUCA radius $R_N$. Furthermore, the expressions in (1b), (1c), and (1d) formulate the uniform circular distribution of elements on each sub-array by defining their azimuth angles $\varphi_{n,k}$ and corresponding Cartesian coordinates $(x_{n,k}, y_{n,k})$. Finally, the condition in (1e) is imposed to mitigate mutual coupling by enforcing a minimum inter-element spacing $D$ between adjacent elements.

Let $x_{n_k}$ denote the modulated signal transmitted from the $k$-th array-element of the $n$-th UCA at the transmitter. It can be expressed as

$$
x_{n_k} = \sum_{l=1-K/2}^{K/2} \frac{1}{\sqrt{K}} s_{n,l} e^{jl\varphi_{n_k}}, \tag{2}
$$

where $l \in [1 - K/2, K/2]$ denotes the spatial mode index of the vortex beam, $s_{n,l}$ represents the data symbol transmitted on the $l$-th orthogonal spatial mode from the $n$-th UCA, and $K$ denotes the number of array-elements in each UCA.

The parameter $\varphi_{n_k}$ denotes the azimuth angle of the $k$-th element on the $n$-th UCA at the transmitter. It is given by

$$
\varphi_{n_k} = \frac{2\pi (k-1)}{K} + n\sigma_t, \tag{3}
$$

where $\sigma_t$ $\left(0 \leq \sigma_t < \frac{2\pi}{N}\right)$ denotes the rotation angle corresponding to the transmit UCA.

Let $\boldsymbol{s}_n = \left[s_{n,1-K/2}, \cdots s_{n,l}, \cdots s_{n,K/2}\right]^T$ be the data symbol vector on the $n$-th UCA. The modulated signal vector $\boldsymbol{X}_n$ on the $n$-th UCA is then expressed as

$$
\boldsymbol{X}_n = [x_0, \cdots, x_k, \cdots, x_{K-1}]^T = \boldsymbol{W}_n \boldsymbol{s}_n, \tag{4}
$$

where $\boldsymbol{W}_n \in \mathbb{C}^{(K) \times (K)}$ denotes the structured beamforming matrix of the $n$-th UCA, which applies the requisite phase shifts to generate orthogonal beams. It is defined as follows:

$$
\boldsymbol{W}_n = \begin{bmatrix} 1 & \cdots & 1 & \cdots & 1 \\ \vdots & \ddots & \vdots & \ddots & \vdots \\ 1 & \cdots & e^{jk\varphi_{n_k}} & \cdots & e^{j\frac{K}{2}\varphi_{n_k}} \\ \vdots & \ddots & \vdots & \ddots & \vdots \\ 1 & \cdots & e^{jk\varphi_{n_K}} & \cdots & e^{j\frac{K}{2}\varphi_{n_K}} \end{bmatrix}. \tag{5}
$$

Assuming the free-space path loss model, the complex channel coefficient between the $k$-th element of the $n$-th transmit UCA and the $v$-th element of the $m$-th receive UCA, denoted by $h_{m_v, n_k}$, is expressed as

$$
h_{m_v, n_k} = \frac{\beta e^{-j\kappa_0 d_{m_v, n_k}}}{2\kappa_0 d_{m_v, n_k}}, \tag{6}
$$

where $d_{m_v, n_k}$ is the Euclidean distance between the corresponding transmit and receive elements, and $\kappa_0 = 2\pi/\lambda$ is the free-space wavenumber. The $d_{m_v, n_k}$ given as follows:

$$
\begin{aligned}
d_{m_v, n_k} &= \sqrt{d^2 + (R_m \cos\theta_{m_v} - R_n \cos\varphi_{n_k})^2 + (R_m \sin\theta_{m_v} - R_n \sin\varphi_{n_k})^2} \\
&= \sqrt{d^2 + R_m^2 + R_n^2 - 2R_m R_n \cos(\theta_{m_v} - \varphi_{n_k})} \\
&= \sqrt{d^2 + R_m^2 + R_n^2} \sqrt{1 - \frac{2R_m R_n \cos(\theta_{m_v} - \varphi_{n_k})}{d^2 + R_m^2 + R_n^2}} \\
&\approx \sqrt{d^2 + R_m^2 + R_n^2} \left(1 - \frac{R_m R_n \cos(\theta_{m_v} - \varphi_{n_k})}{d^2 + R_m^2 + R_n^2}\right) \\
&= \sqrt{d^2 + R_m^2 + R_n^2} - \frac{R_m R_n \cos(\theta_{m_v} - \varphi_{n_k})}{\sqrt{d^2 + R_m^2 + R_n^2}}
\end{aligned} \tag{7}
$$

Substituting Eq. (7) into the channel gain expression, $h_{m_v,n_k}$ can be futher rewrite as

$$h_{m_v,n_k} = \frac{\beta e^{-j\kappa_0\sqrt{d^2+R_m{}^2+R_n{}^2}}}{2\kappa_0 d}e^{j\kappa_0\frac{R_mR_n\cos\left(\theta_{m_v}-\varphi_{n_k}\right)}{\sqrt{d^2+R_m{}^2+R_n{}^2}}}. \quad (8)$$

Consequently, the total signal received at the $v$-th element of the $m$-th receive UCA, denoted by $r_{m,v}$, is the superposition of all incoming signals and can be written as

$$\begin{aligned} r_{mv} &= \sum_{n=1}^{N}\sum_{k=1}^{K}x_{n_k} \\ &= \sum_{n=1}^{N}\sum_{k=1}^{K}\sum_{l=1-K/2}^{K/2}\frac{1}{\sqrt{K}}h_{m_v,n_k}s_{nl}e^{j\varphi_{n_k}l} \\ &= \sum_{n=1}^{N}\sum_{l=1-K/2}^{K/2}s_{nl}h_{m_v,n,l}, \end{aligned} \quad (9)$$

where $h_{m_v,n,l}$ denotes the complex channel gain for the $l$-th spatial mode, as formulated in (10), which involves the Bessel function [8] of the first kind of order $l$, denoted by $J_l(\cdot)$, defined as

$$J_l\left(\alpha\right) = \frac{j^l}{2\pi}\int_0^{2\pi}e^{j\tau l}e^{j\alpha\cos\tau}d\tau. \quad (11)$$

To recover the transmitted spatial modes at the $m$-th receive UCA, a wavefront decomposition matrix is applied to the received spatial samples. Consequently, the recovered signal component corresponding to the spatial mode $l_0$, denoted by $r_{m,l_0}$, is expressed as [24]

$$\begin{aligned} r_{ml_0} &= \sum_{v=1}^{V}r_{mv}e^{-j\left[\frac{2\pi(v-1)}{V}+m\sigma_r\right]l_0} \\ &= \sum_{v=1}^{V}\sum_{n=1}^{N}\sum_{l=1-\frac{K}{2}}^{\frac{K}{2}}s_{nl}h_{m_v,n,l}e^{-j\left[\frac{2\pi(v-1)}{V}+m\sigma_r\right]l_0} \\ &= \sum_{n=1}^{N}\sum_{l=1-\frac{K}{2},l\neq l_0}^{\frac{K}{2}}s_{nl}h_{m_vnl}\sum_{v=1}^{V}e^{j\theta_{m_v}l}e^{-j\left[\frac{2\pi(v-1)}{V}+m\sigma_r\right]l_0} \\ &\quad + \sum_{n=1}^{N}\sum_{v=1}^{V}s_{nl_0}h_{m_vnl_0}e^{j\theta_{m_v}l_0}e^{-j\left[\frac{2\pi(v-1)}{V}+m\sigma_r\right]l_0} \\ &= V\sum_{n=1}^{N}h_{mn,l_0}s_{nl_0}. \end{aligned} \quad (12)$$

The effective channel gain for the $l_0$-th spatial mode between the $n$-th transmit and $m$-th receive UCAs, denoted by $h_{mn,l_0}$, is derived as

$$h_{mn,l_0} = \frac{\beta\sqrt{K}e^{-j\kappa_0\sqrt{d^2+R_m^2+R_n^2}}e^{j\theta_{m_0}l_0}}{2\kappa_0 dj^{l_0}}J_{l_0}\left(\frac{\kappa_0 R_m R_n}{\sqrt{d^2+R_m^2+R_n^2}}\right). \quad (13)$$

This expression reveals that the UCA configuration dictates the effective channel response for the structured vortex beam, which is explicitly characterized by the Bessel function of topological charge $l_0$.

Consequently, the effective channel matrix $\boldsymbol{H}_{\mathrm{C},l}$ encompassing all channels for the $l$-th spatial mode under the CUCA configuration is derived as follows:

$$\boldsymbol{H}_{\mathrm{C},l} = \begin{bmatrix} h_{11,l} & \cdots & h_{1n,l} & \cdots & h_{1N,l} \\ \vdots & \ddots & \vdots & \ddots & \vdots \\ h_{m1,l} & \cdots & h_{mn,l} & \cdots & h_{mN,l} \\ \vdots & \ddots & \vdots & \ddots & \vdots \\ h_{M1,l} & \cdots & h_{Mn,l} & \cdots & h_{MN,l} \end{bmatrix}. \quad (14)$$

Using the zero-forcing detection scheme, the estimate value of transmit signal, denoted by $\tilde{\boldsymbol{s}}_l = [\tilde{s}_{1,l}, \cdots, \tilde{s}_{M,l}]^T$, can be derived as follows:

$$\tilde{\boldsymbol{s}}_l = \boldsymbol{s}_l + \left(\boldsymbol{H}_{\mathrm{C},l}{}^H\boldsymbol{H}_{\mathrm{C},l}\right)^{-1}\boldsymbol{H}_{\mathrm{C},l}{}^H\boldsymbol{z}_l, \quad (15)$$

where $\boldsymbol{z}_l = [z_{1,l}, \cdots, z_{M,l}]^T$ is the received noise vector corresponding to the $l$-th spatial mode.

The achievable SE of the proposed CUCA-based transmission scheme is formulated as

$$C_{\mathrm{CUCA}} = \sum_{l=1-K/2}^{K/2}B\log_2\left[I + \frac{|\boldsymbol{s}_l|^2}{\delta_l{}^2\left|\left(\boldsymbol{H}_{\mathrm{C},l}{}^H\boldsymbol{H}_{\mathrm{C},l}\right)^{-1}\boldsymbol{H}_{\mathrm{C},l}{}^H\right|^2}\right], \quad (16)$$

where $\delta_l{}^2$ denotes the variance of the received noise corresponding to the signal vector $\boldsymbol{z}_l$ for the $l$-th spatial mode.

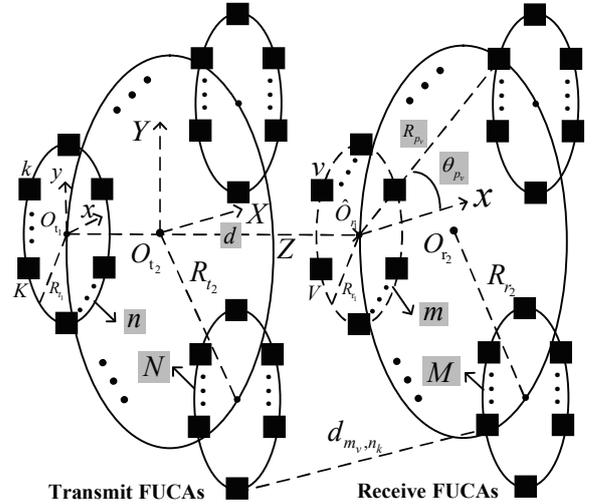

Fig. 3. The geometric configuration of the FUCAs topology.

### B. Structured Beam Generation via Fractal UCAs

Figure 3 illustrates the geometric configuration of the FUCAs topology. As depicted, $N$ transmit UCAs (indexed by $n = 1, \ldots, N$) and $M$ receive UCAs (indexed by $m = 1, \ldots, M$) are uniformly distributed along the primary circumference. Moreover, the constituent antenna elements within each subarray, indexed by $k$ ($1 \le k \le K$) for the transmitter and $v$ ($1 \le v \le V$) for the receiver, are uniformly spaced along the local circumference. The topology is characterized by two hierarchical radii: the primary radius, denoting the distance

$$h_{m_v,n,l} = \sum_{k=1}^{K} \frac{\beta e^{-j\kappa_0\sqrt{d^2+R_m{}^2+R_n{}^2}}}{2\kappa_0 d\sqrt{K}} e^{j\kappa_0 \frac{R_m R_n \cos(\varphi_{n_k}-\theta_{m_v})}{\sqrt{d^2+R_m{}^2+R_n{}^2}}} e^{j\varphi_{n_k} l}$$

$$= \frac{\beta K e^{-j\kappa_0\sqrt{d^2+R_m{}^2+R_n{}^2}} e^{j\theta_{m_v} l}}{2\kappa_0 d\sqrt{K} j^l} \frac{j^l}{K} \sum_{k=1}^{K} e^{j(\varphi_{n_k}-\theta_{m_v})l} e^{j\kappa_0 \frac{R_m R_n \cos(\varphi_{n_k}-\theta_{m_v})}{\sqrt{d^2+R_m{}^2+R_n{}^2}}}$$

$$\approx \frac{\beta\sqrt{K} e^{-j\kappa_0\sqrt{d^2+R_m{}^2+R_n{}^2}} e^{j\theta_{m_v} l}}{2\kappa_0 d j^l} \frac{j^l}{2\pi} \int_0^{2\pi} e^{j(\varphi_{n_k}-\theta_{m_v})l} e^{j\kappa_0 \frac{R_m R_n \cos(\varphi_{n_k}-\theta_{m_v})}{\sqrt{d^2+R_m{}^2+R_n{}^2}}} d(\varphi_{n_k})$$

$$= \frac{\beta\sqrt{K} e^{-j\kappa_0\sqrt{d^2+R_m{}^2+R_n{}^2}} e^{j\theta_{m_v} l}}{2\kappa_0 d j^l} J_l\left(\frac{\kappa_0 R_m R_n}{\sqrt{d^2+R_m{}^2+R_n{}^2}}\right)$$

(10)

from the global origin to the center of each constituent UCA, and the secondary radius, representing the distance from the local center of a sub-array to its antenna elements. These radii are denoted by $R_{t_2}$ and $R_{t_1}$ for the transmitter, and $R_{r_2}$ and $R_{r_1}$ for the receiver, respectively. We denote by $R_{tE}$ and $R_{rE}$ the radii of the entire transmit and receive FUCAs, respectively, where $R_{tE} = R_{t_2} + R_{t_1}$ and $R_{rE} = R_{r_2} + R_{r_1}$. Generally, we assume that $R_{tE} = R_{rE}$. To precisely describe the spatial arrangement, we define two coordinate systems: a local coordinate system ($o$-$xyz$) for each UCA and a global coordinate system ($O$-$XYZ$) for the entire FUCA. In the local system, the $z$-axis is the centering normal line pointing to the directly opposite receive UCA, while the $x$-axis is defined by the vector connecting the UCA center to the first antenna element ($k = 1$ or $v = 1$). Similarly, in the global coordinate system, the $Z$-axis is defined as the surface normal pointing towards the counterpart receive FUCA, while the $X$-axis extends from the global origin to the center of the first constituent UCA ($n = 1$ or $m = 1$). The orientations of the $y$-axis and $Y$-axis are uniquely determined by the right-hand rule. Regarding the angular parameters, we define them across two hierarchical levels. First, at the local level, let $\varphi_{n,k}$ and $\theta_{m,v}$ denote the azimuth angles of the antenna elements within the $n$-th transmit and $m$-th receive UCAs, respectively. Second, at the global level, let $\Phi_n$ and $\Phi_m$ represent the azimuthal positions of the centers of the transmit and receive UCAs on the primary circumference. The angular displacement between them is defined as $\nabla\Phi_{mn} = \Phi_m - \Phi_n$.

The transmit and receive FUCAs are assumed to share identical geometric structures. However, in contrast to the strictly coaxial alignment inherent in CUCA systems, the relative spatial positions of a transmit UCA and a receive UCA within the FUCA architecture can be either coaxial or noncoaxial. To rigorously characterize this complex geometry, we employ the orthogonal projection method illustrated in Fig. 3. Specifically, consider the orthogonal projection of a transmit UCA (centered at $o_{t_1}$) onto the receiver plane. Let $\hat{o}_{r_1}$ denote the center of this projected ring. We subsequently construct a virtual ring centered at $\hat{o}_{r_1}$ with a radius $R_{p_v}$, defined as the Euclidean distance from the projection center $\hat{o}_{r_1}$ to the target array element $p_v$ on the receive UCA. Accordingly, $\theta_{p_v}$ is defined as the azimuth angle formed by the vector connecting $\hat{o}_{r_1}$ to the element $p_v$ relative to the positive $x$-axis.

To ensure the uniform distribution of array elements in FUCA, the array topology must strictly adhere to the specific criteria outlined in (17). Specifically, in the design of FUCAs, the radii $R_n$, azimuth angles $\varphi_{n,k}$, and the number of elements per UCA, denoted by $K$, are required to satisfy predefined geometric constraints. The conditions governing the FUCA topology are formulated as follows:

$$\text{FUCA:} \begin{cases} R_{t_1} < R_{t_2}, R_E = R_{t_1} + R_{t_2}; & (17a) \\ \varphi_{n,k} = \dfrac{2\pi k}{K_n}, \nabla\Phi_{mn} = \dfrac{2\pi(m-n)}{N}; & (17b) \\ x_{n,k} = R_{t_2}\cos\Phi_n + R_{t_1}\cos\varphi_{n,k}; & (17c) \\ y_{n,k} = R_{t_2}\sin\Phi_n + R_{t_1}\sin\varphi_{n,k}; & (17d) \\ \|\boldsymbol{t}_{n,k} - \boldsymbol{t}_{n+1,k+1}\| \gg D. & (17e) \end{cases}$$

Constraint (17a) establishes a hierarchical radial structure, wherein constituent UCAs of radius $R_{t_1}$ are distributed along a primary UCA of radius $R_{t_2}$, with the total aperture radius $R_E$ defined as their sum. Furthermore, the element placement, governed by (17b)–(17d), yields an epicyclic distribution: specifically, elements are uniformly spaced on each constituent UCA, while these UCAs are, in turn, uniformly distributed along the primary FUCA ring. Finally, the condition in (17e) is imposed to mitigate mutual coupling by enforcing a minimum inter-element spacing $D$ between adjacent elements.

Assuming the free-space path loss model, the complex channel coefficient between the $k$-th element of the $n$-th transmit UCA and the $v$-th element of the $m$-th receive UCA, denoted by $h_{m_v,n_k}$, is expressed as

$$h_{m_v,n_k} = \frac{\beta e^{-j\kappa_0 d_{m_v,n_k}}}{2\kappa_0 d_{m_v,n_k}}, \tag{18}$$

where $d_{m_v,n_k}$ is the Euclidean distance between the corresponding transmit and receive elements, and $\kappa_0 = 2\pi/\lambda$ is the free-space wavenumber. The $d_{m_v,n_k}$ given as follows [25]:

$$\begin{aligned} d_{m_v,n_k} &= \sqrt{d^2 + (R_{p_v}\cos\theta_{p_v} - R_{t_1}\cos\varphi_{n_k})^2 + (R_{p_v}\sin\theta_{p_v} - R_{t_1}\sin\varphi_{n_k})^2} \\ &= \sqrt{d^2 + R_{p_v}{}^2 + R_{t_1}{}^2 - 2R_{p_v}R_{t_1}\cos(\theta_{p_v} - \varphi_{n_k})} \\ &= \sqrt{d^2 + R_{p_v}{}^2 + R_{t_1}{}^2}\sqrt{1 - \frac{2R_{p_v}R_{t_1}\cos(\theta_{p_v} - \varphi_{n_k})}{d^2 + R_{t_1}{}^2 + R_{p_v}{}^2}} \\ &\approx \sqrt{d^2 + R_{p_v}{}^2 + R_{t_1}{}^2}\left(1 - \frac{R_{t_1}R_{p_v}\cos(\theta_{p_v} - \varphi_{n_k})}{d^2 + R_{t_1}{}^2 + R_n{}^2}\right) \\ &= \sqrt{d^2 + R_{p_v}{}^2 + R_{t_1}{}^2} - \frac{R_{t_1}R_{p_v}\cos(\theta_{p_v} - \varphi_{n_k})}{\sqrt{d^2 + R_{t_1}{}^2 + R_{p_k}{}^2}}. \end{aligned}$$

(19)

Significantly, the proposed projection framework is generic and encompasses the scenario of strictly aligned UCA pairs. In this special case, the virtual projection parameters degenerate to the intrinsic physical parameters, specifically $R_{p_v} = R_{r_1}$ and $\theta_{p_v} = \theta_{m,v}$. Consequently, the distance $d_{m_v,n_k}$ simplifies to the classical expression for coaxial UCAs, given by

$$
\begin{aligned}
d_{m_v,n_k} &= \sqrt{d^2 + R_{r_1}{}^2 + R_{t_1}{}^2 - 2R_{r_1}R_{t_1}\cos\left(\theta_{m_v} - \varphi_{n_k}\right)} \\
&\approx \sqrt{d^2 + R_{t_1}{}^2 + R_{r_1}{}^2} - \frac{R_{t_1}R_{r_1}\cos\left(\theta_{m_v} - \varphi_{n_k}\right)}{\sqrt{d^2 + R_{t_1}{}^2 + R_{r_1}{}^2}}
\end{aligned}
\tag{20}
$$

It is evident from the established geometry that the virtual radius $R_{p_v}$ and azimuth $\theta_{p_v}$ associated with the receive array element $p_v$ vary distinctively depending on its spatial position. Fig. 4 illustrates the geometric relationship between the virtual radius $R_{p_v}$ and the receive azimuth $\theta_{p_v}$. The expression for $R_{p_v}$ is formulated in Eq. (21).

From the geometry illustrated in Fig. 4, we derive

$$
\begin{cases}
\sin\left(\theta_{p_v} - \varphi_{mn}\right) = \dfrac{R_{r_1}\sin\left(\theta_{m_v} - \varphi_{mn}\right)}{R_{p_v}}; \\[2mm]
\cos\left(\theta_{p_v} - \varphi_{mn}\right) = \dfrac{p_{mn} + R_{r_1}\cos\left(\theta_{m_v} - \varphi_{mn}\right)}{R_{p_v}}.
\end{cases}
\tag{22}
$$

Subsequently, the azimuth angle $\theta_{p_v}$ is expressed as

$$
\theta_{p_v} = \arctan\left(\frac{R_{r_1}\sin\left(\theta_{m_v} - \varphi_{mn}\right)}{p_{mn} + R_{r_1}\cos\left(\theta_{m_v} - \varphi_{mn}\right)}\right) + \varphi_{mn}.
\tag{23}
$$

Consequently, the total signal received by the $v$-th element of the $m$-th receive UCA, denoted by $r_{m,v}$, can be expressed as the linear superposition of signals from all $N$ transmit UCAs

$$
\begin{aligned}
r_{mv} &= \sum_{n=1}^{N}\sum_{k=1}^{K} x_{nk} \\
&= \sum_{n=1}^{N}\sum_{k=1}^{K}\sum_{l=1-K/2}^{K/2}\frac{1}{\sqrt{K}}h_{m_v,n_k}s_{nl}e^{j\varphi_{n_k}l} \\
&= \sum_{n=1}^{N}\sum_{l=1-K/2}^{K/2}s_{nl}h_{m_v,n,l}
\end{aligned}
\tag{24}
$$

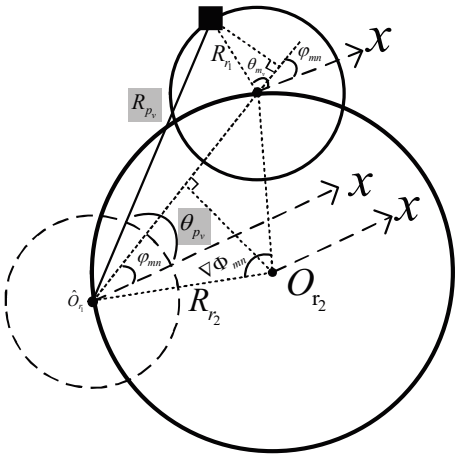

Fig. 4. Geometric illustration of the virtual receive radius and azimuth angle.

To extract the spatial modes at the $m$-th receive UCA, a $V$-point DFT is applied to the received spatial samples. The recovered mode component corresponding to the $l$-th spatial mode, denoted by $r_{m,l}$, is subsequently expressed as

$$
\begin{aligned}
r_{ml} &= \sum_{v=1}^{V}\sum_{n=1}^{N}\sum_{k=1}^{K}\sum_{l=1-K/2}^{K/2}\frac{h_{m_v,n_k}s_{nl}}{\sqrt{K}}e^{j\varphi_{n_k}l}e^{-j\left[\frac{2\pi(v-1)}{V} + m\sigma_r\right]l} \\
&= \begin{cases} V\sum_{n=1}^{N}h_{mn,l_0}s_{nl_0}, & l = l_0; \\[2mm] 0 & , l \neq l_0. \end{cases}
\end{aligned}
\tag{25}
$$

The channel gain for the $l_0$-th spatial mode between the $n$-th transmit and $m$-th receive UCA, denoted by $h_{mn,l_0}$, is derived as

$$
h_{mn,l_0} = \frac{\beta\sqrt{K}e^{-j\kappa_0\sqrt{d^2 + R_{p_v}^2 + R_{t_1}^2}}e^{j\theta_{m_0}l_0}}{2\kappa_0 d j^{l_0}}J_{l_0}\left(\frac{\kappa_0 R_{p_v}R_{t_1}}{\sqrt{d^2 + R_{p_v}^2 + R_{t_1}^2}}\right).
\tag{26}
$$

The gain matrix $\boldsymbol{H}_{\mathrm{F},l}$ of all channels on the FUCA based $l$th mode, can be derived as follows:

$$
\boldsymbol{H}_{\mathrm{F},l} = \begin{bmatrix}
h_{11,l} & \cdots & h_{1n,l} & \cdots & h_{1N,l} \\
\vdots & \ddots & \vdots & \ddots & \vdots \\
h_{m1,l} & \cdots & h_{mn,l} & \cdots & h_{mN,l} \\
\vdots & \ddots & \vdots & \ddots & \vdots \\
h_{M1,l} & \cdots & h_{Mn,l} & \cdots & h_{MN,l}
\end{bmatrix}.
\tag{27}
$$

The SE of the proposed FUCA-based transmission scheme is formulated as

$$
C_{\mathrm{FUCA}} = \sum_{l=1-K/2}^{K/2}B\log_2\left[I + \frac{|\boldsymbol{s}_l|^2}{\delta_l{}^2\left|\left(\boldsymbol{H}_{\mathrm{F},l}{}^H\boldsymbol{H}_{\mathrm{F},l}\right)^{-1}\boldsymbol{H}_{\mathrm{F},l}{}^H\right|^2}\right],
\tag{28}
$$

where $\delta_l{}^2$ denotes the variance of the received noise corresponding to the signal vector $\boldsymbol{z}_l$ for the $l$-th spatial mode.

### C. Joint Optimization of Array Topology and Structured Beamforming

The primary objective is to maximize the achievable SE of the proposed MA-enabled reconfigurable array architecture for structured beam transmission. Accordingly, the capacity maximization problem, denoted by $\boldsymbol{P}_1$, is formulated as

$$
\boldsymbol{P}_1: \max_{\{\boldsymbol{t}_n\},\{\boldsymbol{r}_n\},\{\boldsymbol{W}_n\}} C^* \tag{29a}
$$

$$
\text{s.t.} \quad \sum_{n=1}^{N}\|\boldsymbol{W}_n\|^2 \leq P_{\max}, \tag{29b}
$$

$$
\boldsymbol{t}_n \in S_T, \quad \boldsymbol{r}_n \in S_R, \tag{29c}
$$

Topology constraints in Eq. (1) or Eq. (17), (29d)

where $C^*$ represents the capacity for either the CUCA or FUCA topology. Constraint (29b) imposes the total transmit power budget $P_{\max}$. Constraints (29c) and (29d) guarantee that the MA positions satisfy the discrete search space requirements $(S_T, S_R)$ and the topology-specific geometric relationships, respectively.

$$R_{p_v} = \left\{ \left[ R_{r_1} \cos\left(\theta_{m_v} - \varphi_{mn}\right) + 2R_{r_2} \sin\nabla\Phi_{mn} \right]^2 + R_{r_1}{}^2 \sin^2\left(\theta_{m_v} - \varphi_{mn}\right) \right\}^{\frac{1}{2}}$$
$$= \left\{ R_{r_1}{}^2 \cos^2\left(\theta_{m_v} - \varphi_{mn}\right) + 4R_{r_2}{}^2 \sin^2\nabla\Phi_{mn} + 4R_{r_2}R_{r_1}\sin\nabla\Phi_{mn}\cos\left(\theta_{m_v} - \varphi_{mn}\right) + R_{r_1}{}^2\sin^2\left(\theta_{m_v} - \varphi_{mn}\right) \right\}^{\frac{1}{2}} \quad (21)$$
$$= \left\{ 1 + 4R_{r_2}\sin\nabla\Phi_{mn}\left(R_{r_2}\sin\nabla\Phi_{mn} + R_{r_1}\cos\left(\theta_{m_v} - \varphi_{mn}\right)\right) \right\}^{\frac{1}{2}}$$

---

**Algorithm 1** Alternating Optimization for Solving $\boldsymbol{P}_1$

---

1: **Input:** $N_r, R_N, S_T, S_R, P_{\max}, \varepsilon$
2: **Output:** Optimal topology $\mathscr{G}^*$, topology parameters $\mathscr{T}^* = \{N^*, \{K_n^*\}, \{R_n^*\}\}, \{\boldsymbol{t}_n{}^*\}, \{\boldsymbol{r}_n{}^*\}, \{\boldsymbol{W}_n{}^*\}, C^*$
3: **for** $\mathscr{G} \in \{\text{CUCA}, \text{FUCA}\}$ **do**
4:    Initialize feasible topology $\mathscr{T}^{(0)} = \{N^{(0)}, \{K_n^{(0)}\}, \{R_n^{(0)}\}\}$, $C^{(0)} \leftarrow C(\mathscr{G}, \mathscr{T}^{(0)})$, $i \leftarrow 0$
5:    **while** true **do**
6:       Generate candidate set $\mathcal{S}_i$ satisfying Eq. (1) if $\mathscr{G} = $ CUCA, or Eq. (17) if $\mathscr{G} = $ FUCA
7:       **if** $\mathcal{S}_i = \emptyset$ **then** break
8:       **end if**
9:       Optimize $\{R_n\}$ for all $\mathscr{T}_s \in \mathcal{S}_i$ with $R_n \leq R_N$, and compute $C_s$
10:      $(\mathscr{T}^{(i+1)}, C^{(i+1)}) \leftarrow \arg\max_{\mathscr{T}_s \in \mathcal{S}_i} C_s$
11:      **if** $|C^{(i+1)} - C^{(i)}| < \varepsilon$ **then** break
12:      **end if** $\quad i \leftarrow i + 1$
13:    **end while**
14:    **if** $C^{(i+1)} > C^*$ **then**
15:      $\mathscr{G}^* \leftarrow \mathscr{G}, \mathscr{T}^* \leftarrow \mathscr{T}^{(i+1)}, C^* \leftarrow C^{(i+1)}$
16:    **end if**
17: **end for**
18: Compute $\{\boldsymbol{W}_n{}^*\}$, $\{\boldsymbol{r}_n{}^*\}$, and $\{\boldsymbol{t}_n{}^*\}$ from $\mathscr{G}^*$ and $\mathscr{T}^*$
19: **return** $\mathscr{G}^*, \mathscr{T}^*, \{\boldsymbol{t}_n{}^*\}, \{\boldsymbol{r}_n{}^*\}, \{\boldsymbol{W}_n{}^*\}$, and $C^*$

---

To address problem $\boldsymbol{P}_1$, we propose an alternating optimization framework that jointly configures the array topology and determines the corresponding antenna element positions. The proposed algorithm is designed to traverse distinct topology families, iteratively refining both discrete topology parameters and continuous geometric parameters to maximize the achievable channel capacity. Specifically, we focus on two candidate topology families: CUCA and FUCA. For each family, the process initiates by constructing a feasible initial topology. Subsequently, a candidate-based iterative search is executed, wherein multiple feasible topological variants are generated and evaluated in each iteration. The candidate yielding the highest capacity gain is retained as the seed for the next iteration. The procedure terminates when the capacity improvement saturates or when the search space of feasible candidates is exhausted.

As outlined in **Algorithm 1**, the outer loop traverses the topology categories, denoted by $\mathscr{G} \in \{\text{CUCA}, \text{FUCA}\}$. In each iteration, a candidate topology set $\mathcal{S}_i$ is generated in accordance with the topology-specific geometric constraints, specified in Eq. (1) for CUCA and Eq. (17) for FUCA. Subsequently, for each candidate configuration within $\mathcal{S}_i$, the continuous radius parameters are optimized within the feasible search spaces $S_T$ and $S_R$, subject to the maximum aperture constraint $R_n \leq R_N$. Subsequently, the corresponding channel capacity is evaluated. Notably, the proposed algorithm ensures a non-decreasing objective value across iterations, as the topology update is strictly selected from candidates that maximize the achievable capacity. Furthermore, given that the set of feasible topologies under the specified constraints possesses a finite cardinality, the algorithm is theoretically guaranteed to converge within a finite number of iterations. Despite the involvement of both discrete and continuous optimization variables, the computational complexity remains tractable. This is attributed to the localized candidate generation strategy and the low dimensionality of the continuous optimization regarding the radius parameters. Consequently, the proposed framework is highly viable for practical array topology design under realistic geometric constraints.

Distinct from traditional plane waves, structured beams possess unique helical phase wavefronts, where both the beam spatial structure and spatial mode significantly govern communication performance. Crucially, the number of array elements fundamentally determines the array geometry, which in turn dictates the efficacy of the corresponding structured beam synthesis. However, it is imperative to recognize that merely increasing the element count does not invariably yield superior performance under specific geometric configurations. To address this, we introduce a flexible element allocation strategy, allowing the number of active array elements to be adaptively adjusted within a maximum budget $N_r$ to accommodate diverse array geometries. In this paper, we adopt a symmetric system configuration where identical array geometries are employed at both the transmitter and receiver. This implies that the MAs in the transmit and receive arrays occupy spatially corresponding positions. This assumption aligns with the majority of existing structured beam systems. While architectures with mismatched transmitter-receiver topologies are theoretically possible, they necessitate highly sophisticated signal processing algorithms for wavefront decomposition, which falls beyond the scope of this work. Consequently, the optimization complexity in **Algorithm 1** is significantly reduced, as it suffices to optimize the antenna positions for one side and map the configuration to the other.

## IV. RECONFIGURABLE ARRAY TOPOLOGIES

Building upon the fundamental CUCA and FUCA architectures, we present a diverse set of array topologies. Fig. 5 and Fig. 6 illustrate the configurations for 16-element and 32-element arrays, respectively. The proposed framework enables dynamic reconfiguration, allowing for mutual transformation between any arbitrary pair of these topologies. Furthermore, as the number of array elements increases, the design space expands, allowing for the synthesis of an even richer variety of reconfigurable array topologies. Specifically, a unified hardware platform is utilized to adapt to diverse communication scenarios, obviating the need for physical antenna replacement. By leveraging software-defined reconfiguration, versatile array topologies and beam structures can be synthesized to align with specific communication scenarios [26], [27]. Distinct from existing MA-based communication systems, this paper proposes novel array models utilizing the CUCA and FUCA architectures. In the proposed framework, the spatial positions of the antenna elements are jointly governed by the UCA radius, the number of constituent elements, and the rotation angle, thereby yielding distinct geometric patterns. Fundamentally, the array topology, representing the collective spatial arrangement of the antenna elements, enables diverse structural configurations achieved through varied element layouts.

Conventional antenna arrays utilizing FPAs are inherently constrained by their rigid topological structures. Specifically, once the array topology is established, the corresponding OAM modulation vector becomes invariant. Consequently, neither the array topology nor the transmission scheme can be flexibly reconfigured to adapt to complex and dynamic wireless communication scenarios. In contrast to traditional FPAs-based arrays, this paper introduces the paradigm of MAs, which significantly expands the topological degrees of freedom. This flexibility enables the synthesis of a diverse range of dynamically reconfigurable array topologies, encompassing UCAs, concentric circular arrays, fractal arrays, quasi-fractal arrays, linear arrays, uniform planar square arrays, spiral arrays, and radial line arrays [28], [29]. Moreover, these diverse array geometries demonstrate that the structured beamforming vector is not rigidly coupled to the element positions. Even with an identical geometric layout, the structured beamforming vector can be varied, indicating that there exists no direct one-to-one correspondence between the physical element positions and the modulation vectors. Inherently, structured beam-based wireless communication inherently involves the joint design of array topology and modulation schemes. Notably, the system can switch between different modulation schemes without altering the array geometry. This capability is not limited to transmitting structured beams, but also extends to transmitting conventional plane-wave beams.

Furthermore, the system can dynamically provide diverse array geometries alongside their corresponding beamforming vectors. This joint reconfigurability allows the system to adapt to a wide range of wireless communication scenarios and satisfy varying performance requirements. In addition to the CUCA and FUCA-based topologies proposed in this work, we reference our prior study [8], which introduced the Quasi-Fractal UCA (QF-UCA) architecture [30]. While the intricate design details and specific beam generation schemes are comprehensively covered in [8] and thus omitted here for brevity, the resulting QF-UCA array topology is illustrated in Fig. 5(h) for comparative purposes. The total element displacement required to transition between two array topologies, denoted by $D\left(G_A, G_B\right)$, is defined as the minimum aggregate Euclidean distance traversed by the array elements during the reconfiguration process [31]. To ensure a fair comparison of the reconfiguration overhead, all considered array architectures are normalized to share an identical maximum aperture radius.

$$D\left(G_A, G_B\right) = \min_{\pi \in \Pi} \sum_{i=1}^{N_r} \left\| \mathbf{p}_i^{(G_A)} - \mathbf{p}_{\pi(i)}^{(G_B)} \right\|_2, \qquad (30)$$

where $\Pi$ denotes the set of all possible topologies mapping the element indices of $G_A$ to those of $G_B$.

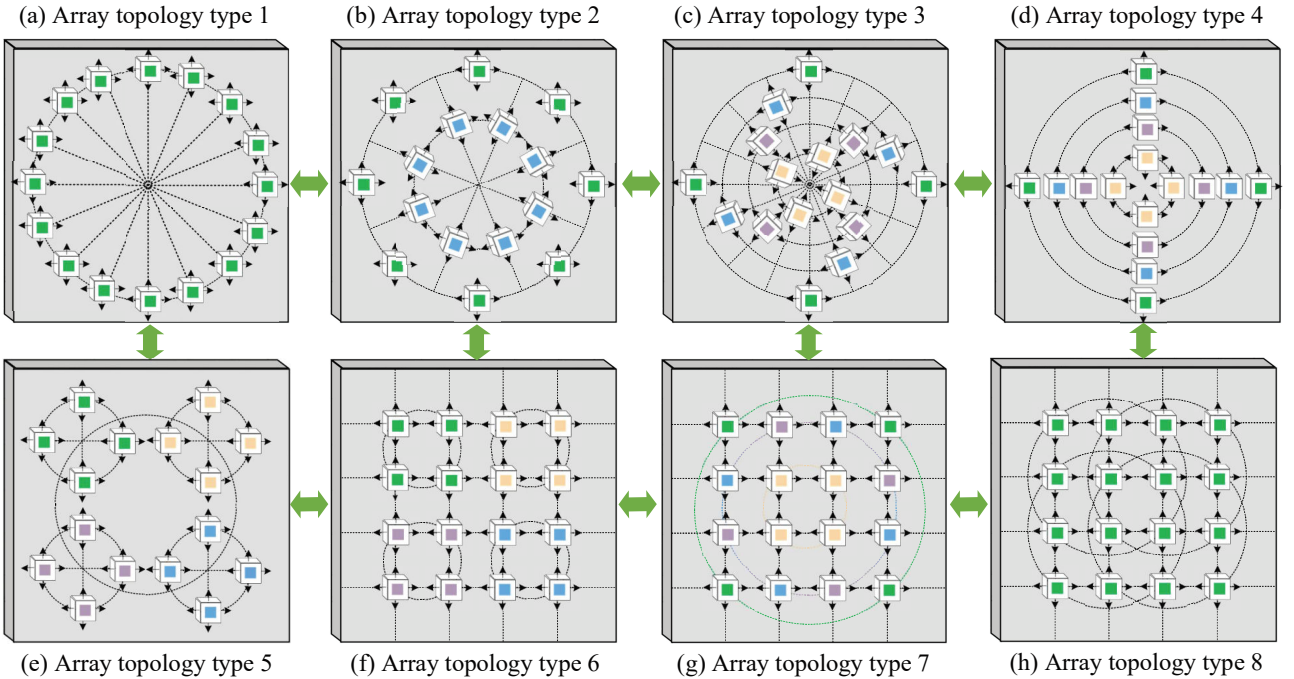

(a) Array topology type 1    (b) Array topology type 2    (c) Array topology type 3    (d) Array topology type 4

(e) Array topology type 5    (f) Array topology type 6    (g) Array topology type 7    (h) Array topology type 8

Fig. 5. The different array topologies with 16 elements.

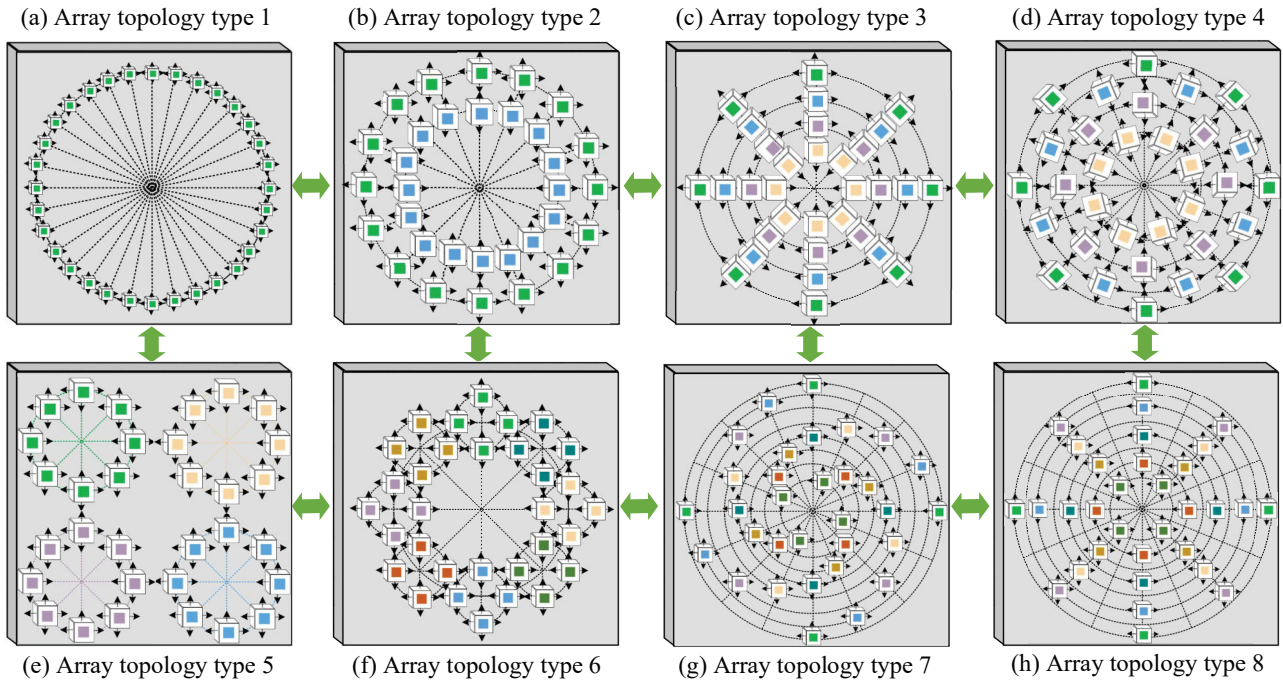

(a) Array topology type 1   (b) Array topology type 2   (c) Array topology type 3   (d) Array topology type 4

(e) Array topology type 5   (f) Array topology type 6   (g) Array topology type 7   (h) Array topology type 8

Fig. 6. The different array topologies with 32 elements.

## V. PERFORMANCE EVALUATIONS

In this section, numerical simulation results are presented to evaluate the performance of the proposed MA-enabled reconfigurable array architecture for structured beam transmission. Without loss of generality, we focus on spatial mode multiplexing exemplified by vortex beams carrying OAM to validate the proposed framework. The simulation parameters are summarized in Table II unless specified otherwise.

TABLE II
SIMULATION PARAMETERS

| $\beta$ | $4\pi$ | $f$ | 5.8 GHz |
|---|---|---|---|
| $R_{rE}$ | 2 m | $R_{tE}$ | 2 m |
| $\sigma^2$ | 10 mW | $P_{\max}$ | 1 W |
| $d$ | 100 m | $B$ | 10 MHz |

Figure 7 shows the SEs achieved by four switchable 16-element array topologies: a UCA with 16 elements, a CUCA with 2 circles, each containing 8 elements (2 rings × 8 elements), a CUCA with 4 circles, each containing 4 elements (4 rings × 4 elements), and a FUCA with 4 circles, each containing 4 elements (4 rings × 4 elements), all under identical system settings. The corresponding topologies are shown in Figs. 5(a), 5(b), 5(c), and 5(e). The radius is fixed at $R_E = 2$ m. In the CUCA-based layouts, the ring radii are uniformly reduced from the outermost ring to the center, whereas the FUCA employs two characteristic radii $R_2 = 1.2$ m (the distance from the array center to each sub-UCA center) and $R_1 = 0.8$ m (the radius of each sub-UCA), ensuring that the overall aperture satisfies $R_E = R_1 + R_2$. As observed, the CUCA (4 rings × 4 elements) consistently provides the highest SE across the entire SNR range. This improvement can be attributed to two factors: i) more effective multiplexing of low-order OAM modes, which experience lower propagation loss; and ii) a more uniform utilization of the available aperture enabled by the evenly spaced ring radii, which improves the spatial separability among OAM channels. Under the given parameter settings, the CUCA (2 rings × 8 elements) slightly outperforms the FUCA, while the single-ring UCA achieves the lowest SE. The performance degradation of the UCA is primarily due to its inefficient use of the array aperture, as it relies on higher-order OAM modes to maintain the same number of streams. These higher-order modes suffer from greater attenuation, thus limiting the achievable SE.

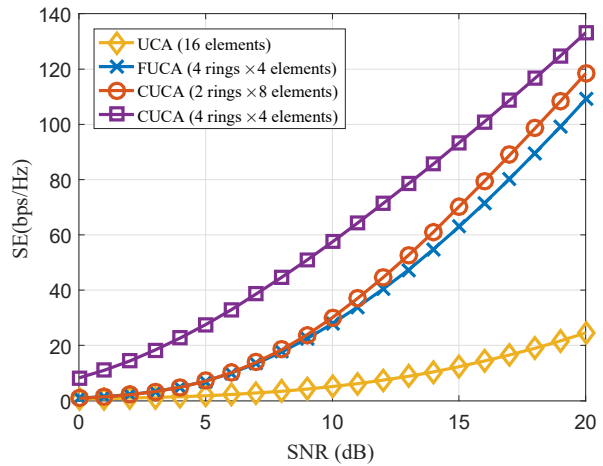

Fig. 7. The SEs of the UCA, CUCA, and FUCA based OAM transmission with 16 array-elements (Figs. 5(a), 5(b), 5(c), and 5(e) are the corresponding array-element layouts, respectively).

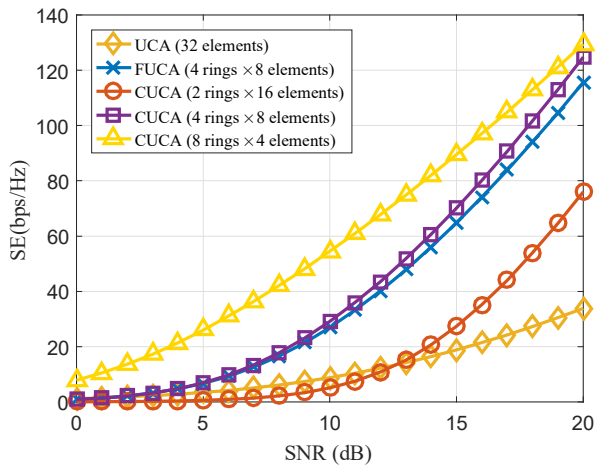

Fig. 8. The SEs of the UCA, CUCA, and FUCA based OAM transmission with 32 array-elements (Figs. 6(a), 6(b), 6(c), 6(g), and 6(e) are the corresponding array-element layouts, respectively).

Figure 8 illustrates the SEs achieved by four switchable 32-element array topologies: a UCA with 32 elements, a CUCA with 2 circles, each containing 16 elements (2 rings × 16 elements), a CUCA with 4 circles, each containing 8 elements (4 rings × 8 elements), a CUCA with 8 circles, each containing 4 elements (8 rings × 4 elements), and a FUCA with 4 circles, each containing 8 elements (4 rings × 8 elements), all under identical system settings. The corresponding topologies are shown in Figs. 6(a), 6(b), 6(c), 6(g), and 6(e). It is observed that the CUCA (8 rings × 4 elements) achieves the highest SE across the entire SNR range, followed by the CUCA (4 rings × 8 elements). This trend suggests that, for a fixed number of array elements, increasing the number of concentric rings while reducing the number of elements per ring is advantageous for enhancing SE. The performance improvement can be attributed to two main factors. On the one hand, the additional concentric rings facilitate more multiplexing reuse of lower-order OAM modes, which experience lower propagation loss. On the other hand, the evenly distributed ring radii enable the array to better utilize the available aperture, thereby enhancing the orthogonality of spatial modes. In contrast, the FUCA (4 rings × 8 elements) exhibits slightly inferior performance compared to the CUCA with the same number of rings. The CUCA with 2 rings and 16 elements shows further SE degradation, while the single-ring UCA yields the lowest SE. This is because the UCA relies more heavily on higher-order OAM modes to support multiplexing, which suffer from more severe attenuation and thus limit the achievable SE.

Figure 9 illustrates the SEs of various array topologies at different transmission distances and array apertures. It is observed that the SE is jointly affected by the array radius $R_E$ and the transmission distance $D$, which together determine the effective channel gain. Specifically, the SE surface exhibits oscillatory behavior with respect to the array radius, resembling

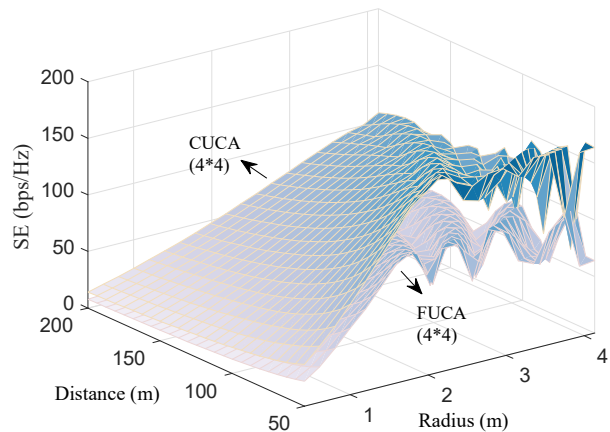

(a) CUCA (4 rings × 4 elements) and FUCA (4 rings × 4 elements).

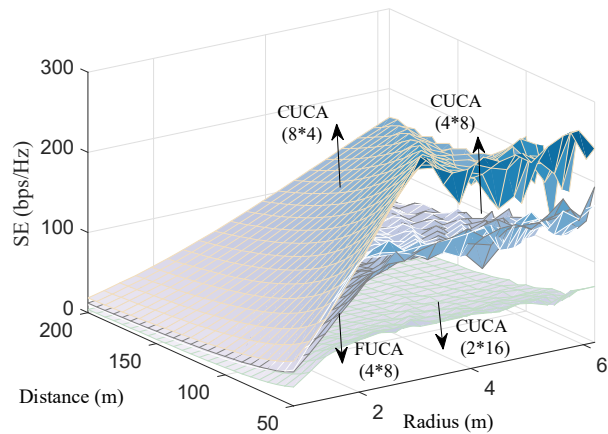

(b) CUCA (8 rings × 4 elements), CUCA (4 rings × 8 elements), FUCA (4 rings × 8 elements), and CUCA (2 rings × 16 elements).

Fig. 9. The SEs of various array topologies at different transmission distances and array apertures.

the characteristics of Bessel functions. This behavior aligns with the analytical channel gain expressions in Eqs. (13) and (26), which highlight the dependence of the received OAM mode power on both the array aperture and the propagation distance. As a result, the SE does not increase monotonically with the array radius, but instead exhibits multiple local maxima. Moreover, different array topologies correspond to distinct optimal radius values. By appropriately adjusting the element positions and selecting an optimal array radius, significant performance gains can be achieved. This observation further demonstrates the advantage of flexible array topology design, where the array geometry can be adapted to the transmission distance to maximize spectral efficiency. Therefore, jointly optimizing the array radius and topology according to the transmission distance provides an effective means to fully exploit the potential of OAM-based communications.

Figure 10 depicts the bit error rate (BER) performance of various array topologies with 32 array elements. As shown in Fig. 10, the CUCA with 8 circles, each containing 4 elements (8 rings × 4 elements) achieves the lowest BER over the entire SNR range, followed by the CUCA with 4 circles, each containing 8 elements (4 rings × 8 elements). This trend is consistent with the SE results shown in Fig. 8, indicating that increasing the number of concentric rings benefits not only the enhancement of spectral efficiency but also the improvement of detection reliability. The superior BER performance can be attributed to a more effective utilization of low-order OAM modes. In contrast, the FUCA (4 rings × 8 elements) exhibits inferior BER performance compared to the CUCA with the same number of rings, due to the nonuniform distribution of array elements within the aperture. The CUCA with 2 rings and 16 elements experiences further performance degradation, while the single-ring UCA yields the poorest BER performance. This is because the UCA relies more heavily on higher-order OAM modes for multiplexing, which experience greater attenuation and lead to increased error rates. The observed performance gain is primarily attributed to the combined design of the array topology and the associated modulation vectors. This joint optimization enhances the effective channel gain and improves the received signal quality, thereby enabling more reliable communication.

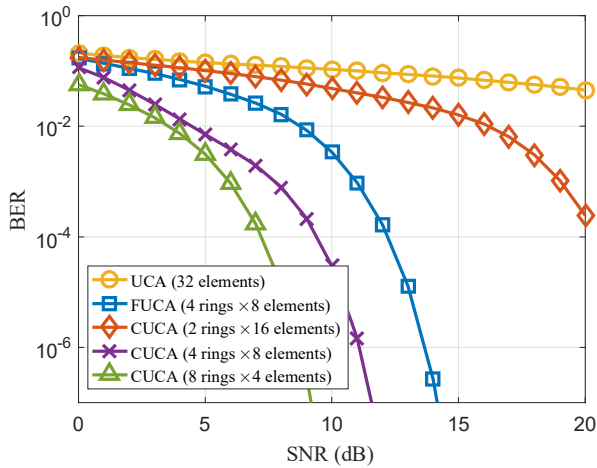

Fig. 10. BER performance of various array topologies with 32 array elements.

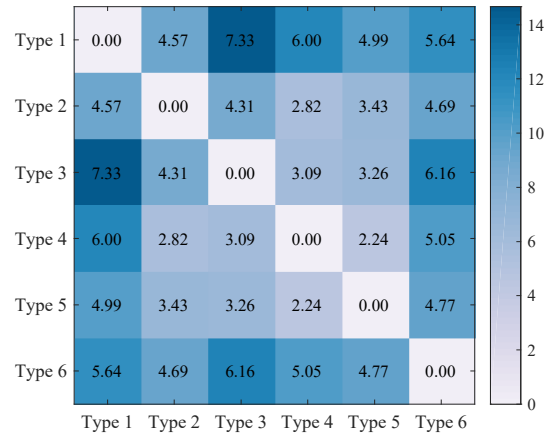

Fig. 11. The total displacement distance of all array-elements for the conversion among different array topologies (The corresponding array topologies are shown in Figs. 5(a), 5(b), 5(c), 5(d), 5(e), and 5(f)).

Figure 11 illustrates the array-element layouts of different topologies and the corresponding element displacement distance required for topology switching, where the distance is evaluated as the total mechanical movement of all array elements. Since the mechanical movement of reconfigurable antennas incurs additional power consumption, the switching cost between different array topologies constitutes an important design factor in practical communication systems. The heatmap illustrates the topology switching distance for the considered 16-element configurations. It can be observed that switching from the conventional UCA to other topologies generally requires a relatively large element displacement, corresponding to the outer region of the heatmap. In contrast, the switching distances among different CUCA configurations with varying numbers of rings, as well as between CUCA and FUCA, are significantly smaller, which are mainly located in the inner region of the heatmap. More importantly, a relatively small amount of element displacement is sufficient to achieve a significant performance improvement. This observation highlights the potential of MAs, where flexible topology reconfiguration enables substantial performance gains with moderate mechanical cost. By allowing the array topology to adapt to different operating scenarios, movable antenna systems introduce an additional degree of freedom in wireless communications, leading to enhanced system performance across diverse application requirements.

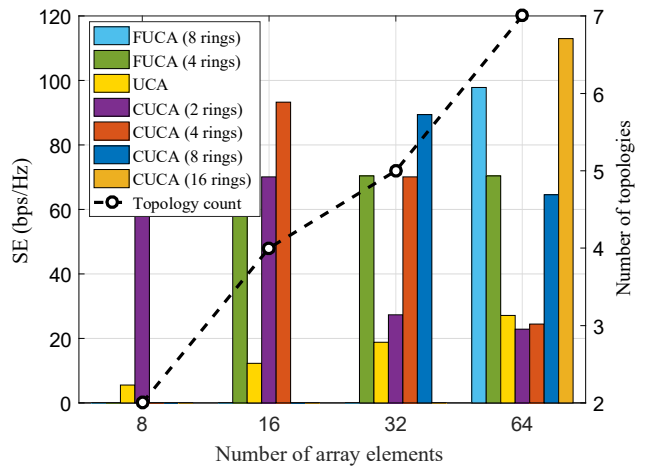

Fig. 12. The SEs of different topologies with varying numbers of elements.

Figure 12 illustrates the SE performance of different antenna topologies versus the number of array elements, together with the corresponding number of available topologies. It is observed that as the number of array elements increases, the number of feasible topologies grows accordingly, enabling the system to switch among a larger set of antenna configurations. With the continuous evolution of wireless communication technologies, the scale of antenna arrays continues to grow. The adoption of MAs introduces flexible and reconfigurable topologies, significantly expanding the design space for antenna deployment. Under varying scenarios and system param-eters, multiple topology options can be supported, enabling the system to select the most favorable topology based on specific performance requirements. Moreover, as the MAs can follow more diverse array movement trajectories, the corresponding shortest-path planning problem becomes increasingly complex. This highlights the necessity of efficient topology selection and movement planning strategies to fully exploit the performance gains enabled by antenna mobility, while managing the associated reconfiguration overhead.

## VI. Conclusions

In this paper, we investigated a novel structured beam communication framework enabled by movable antennas, featuring reconfigurable array topologies to fully exploit the array aperture and ensure adaptability to diverse communication scenarios. Specifically, by leveraging the mobility DoFs of MAs, we established a geometric modeling framework capable of constructing diverse array configurations-including concentric uniform circular arrays, fractal uniform circular arrays, uniform planar square arrays, radially uniform linear arrays, and spiral arrays-utilizing a unified hardware platform. To maximize array aperture efficiency and SE, we developed a joint optimization scheme for the array topology and the structured beamforming vector, supported by rigorous analytical channel gains for orthogonal spatial modes and an efficient alternating optimization algorithm. Numerical results demonstrate that the proposed joint design significantly enhances the SE compared to conventional FPA schemes. Consequently, this work reveals that the synergy between spatially structured beams and the geometric flexibility of MAs offers a reconfigurable and practical solution for high-capacity 6G wireless communications.

## References

[1] C. You, Y. Cai, Y. Liu, M. Di Renzo, T. M. Duman, A. Yener, and A. Lee Swindlehurst, "Next generation advanced transceiver technologies for 6G and beyond," *IEEE Journal on Selected Areas in Communications*, vol. 43, no. 3, pp. 582–627, 2025.

[2] X. Zhang, S. Zheng, X. Xiong, Z. Zhu, Y. Chen, Z. Wang, X. Hui, W. E. I. Sha, and X. Yu, "Structured beamforming based on orbital angular momentum mode-group," *Journal of Lightwave Technology*, vol. 41, no. 7, pp. 1997–2006, 2023.

[3] W. Cheng, W. Zhang, H. Jing, S. Gao, and H. Zhang, "Orbital angular momentum for wireless communications," *IEEE Wireless Communications*, vol. 26, no. 1, pp. 100–107, Feb. 2019.

[4] P. Mollahosseini and Y. Ghasempour, "Fast vortex beam alignment for OAM mode multiplexing in LOS MIMO networks," *IEEE Transactions on Wireless Communications*, vol. 25, pp. 10 546–10 559, 2026.

[5] R. Chen, H. Zhou, M. Moretti, X. Wang, and J. Li, "Orbital angular momentum waves: Generation, detection, and emerging applications," *IEEE Communications Surveys & Tutorials*, vol. 22, no. 2, pp. 840–868, 2020.

[6] R. Lyu, W. Cheng, Q. Du, and T. Q. S. Quek, "Capacity analysis on OAM-based wireless communications: An electromagnetic information theory perspective," *IEEE Transactions on Wireless Communications*, vol. 25, pp. 2528–2543, 2026.

[7] Z. Wang, J. Zhang, B. Xu, W. Yi, E. Björnson, and B. Ai, "Flexible MIMO for future wireless communications: Which flexibilities are possible?" *IEEE Wireless Communications*, pp. 1–10, 2026.

[8] H. Jin, W. Cheng, H. Jing, J Wang, and W. Zhang, "Achieving high capacity transmission with N-dimensional quasi-fractal UCA," *IEEE Transactions on Communications*, vol. 73, no. 11, pp. 11 947–11 963, 2025.

[9] A. Irshad, A. Kosasih, V. Petrov, and E. Björnson, "Pre-optimized irregular arrays versus movable antennas in multi-user MIMO systems," *IEEE Wireless Communications Letters*, vol. 14, no. 8, pp. 2656–2660, 2025.

[10] W. Shi, S. A. Vorobyov, and Y. Li, "ULA fitting for sparse array design," *IEEE Transactions on Signal Processing*, vol. 69, pp. 6431–6447, 2021.

[11] H. Jin, W. Cheng, J. Wang, Q. Du, and W. Zhang, "Achieving high-capacity OAM communication with fluid-antenna-based continuous-aperture arrays," *IEEE Journal on Selected Areas in Communications*, pp. 1–1, 2025.

[12] L. Zhu, W. Ma, W. Mei, Y. Zeng, Q. Wu, B. Ning, Z. Xiao, X. Shao, J. Zhang, and R. Zhang, "A tutorial on movable antennas for wireless networks," *IEEE Communications Surveys & Tutorials*, pp. 1–1, 2025.

[13] L. Zhu, W. Ma, and R. Zhang, "Movable antennas for wireless communication: Opportunities and challenges," *IEEE Communications Magazine*, vol. 62, no. 6, pp. 114–120, 2024.

[14] W. Ma, L. Zhu, and R. Zhang, "MIMO capacity characterization for movable antenna systems," *IEEE Transactions on Wireless Communications*, vol. 23, no. 4, pp. 3392–3407, 2024.

[15] B. Ning, S. Yang, Y. Wu, P. Wang, W. Mei, C. Yuen, and E. Bjornson, "Movable antenna-enhanced wireless communications: General architectures and implementation methods," *IEEE Wireless Communications*, pp. 1–9, 2025.

[16] L. Zhu, W. Ma, and R. Zhang, "Modeling and performance analysis for movable antenna enabled wireless communications," *IEEE Transactions on Wireless Communications*, vol. 23, no. 6, pp. 6234–6250, 2024.

[17] Y. Gao, Q. Wu, and W. Chen, "Joint transmitter and receiver design for movable antenna enhanced multicast communications," *IEEE Transactions on Wireless Communications*, vol. 23, no. 12, pp. 18 186–18 200, 2024.

[18] Z. Dong, Z. Zhou, Z. Xiao, C. Zhang, X. Li, H. Min, Y. Zeng, S. Jin, and R. Zhang, "Movable antenna for wireless communications: Prototyping and experimental results," *IEEE Transactions on Wireless Communications*, vol. 25, pp. 6586–6599, 2026.

[19] F. Qin, J. Bi, J. Ma, C. Gu, H. Zhang, W. Cheng, and S. Gao, "A high-efficiency reconfigurable bidirectional array antenna based on transmit-reflect switchable metasurface," *IEEE Transactions on Antennas and Propagation*, vol. 73, no. 9, pp. 7051–7056, 2025.

[20] J. Zheng, J. Zhang, H. Du, D. Niyato, S. Sun, B. Ai, and K. B. Letaief, "Flexible-position MIMO for wireless communications: Fundamentals, challenges, and future directions," *IEEE Communications*, vol. 31, no. 6, pp. 18–26, 2024.

[21] A. C. Fikes, M. Gal-Katziri, O. S. Mizrahi, D. E. Williams, and A. Hajimiri, "Frontiers in flexible and shape-changing arrays," *IEEE Journal of Microwaves*, vol. 3, no. 1, pp. 349–367, 2023.

[22] R. Cohen and Y. C. Eldar, "Sparse array design via fractal geometries," *IEEE Transactions on Signal Processing*, vol. 68, pp. 4797–4812, 2020.

[23] D. Wang, W. Mei, B. Ning, Z. Chen, and R. Zhang, "Movable antenna enhanced wide-beam coverage: Joint antenna position and beamforming optimization," *IEEE Transactions on Wireless Communications*, vol. 25, pp. 3541–3558, 2026.

[24] H. Jing, W. Cheng, Z. Li, and H. Zhang, "Concentric UCAs based low-order OAM for high capacity in radio vortex wireless communications," *Journal of Communications and Information Networks*, vol. 3, no. 4, pp. 85–100, 2018.

[25] W. Cheng, H. Zhang, L. Liang, H. Jing, and Z. Li, "Orbital-angular-momentum embedded massive MIMO: Achieving multiplicative spectrum-efficiency for mmwave communications," *IEEE Access*, vol. 6, pp. 2732–2745, 2018.

[26] M. Temiz, E. Alsusa, L. Danoon, and Y. Zhang, "On the impact of antenna array geometry on indoor wideband massive MIMO networks," *IEEE Transactions on Antennas and Propagation*, vol. 69, no. 1, pp. 406–416, 2021.

[27] A. M. A. Shaalan, J. Du, and Y.-H. Tu, "Dilated nested arrays with more degrees of freedom (DOFs) and less mutual coupling-part I: The fundamental geometry," *IEEE Transactions on Signal Processing*, vol. 70, pp. 2518–2531, 2022.

[28] H. Jing, W. Cheng, and W. Zhang, "Breaking limits of line-of-sight MIMO capacity in 6G wireless communications," *IEEE Wireless Communications*, vol. 31, no. 5, pp. 28–33, 2024.

[29] A Manikas, A. Sleiman, and I. Dacos, "Manifold studies of nonlinear antenna array geometries," *IEEE Transactions on Signal Processing*, vol. 49, no. 3, pp. 497–506, 2001.

[30] W. Cheng, H. Jing, W. Zhang, K. Zhang, and H. Zhang, "Quasi-fractal UCA-based OAM for highly efficient orthogonal transmission," *IEEE Transactions on Wireless Communications*, vol. 23, no. 11, pp. 16 513–16 526, 2024.

[31] Q. Li, W. Mei, R. Zhang, and B. Ning, "Trajectory optimization for minimizing movement delay in movable antenna systems," *IEEE Transactions on Wireless Communications*, vol. 25, pp. 6986–6999, 2026.